\documentclass[12pt]{article}
\usepackage{graphicx}
\begin{document}


\def\a{\alpha}
\def\b{\beta}
\def\c{\varepsilon}
\def\d{\delta}
\def\e{\epsilon}
\def\f{\phi}
\def\g{\gamma}
\def\h{\theta}
\def\k{\kappa}
\def\l{\lambda}
\def\m{\mu}
\def\n{\nu}
\def\p{\psi}
\def\q{\partial}
\def\r{\rho}
\def\s{\sigma}
\def\t{\tau}
\def\u{\upsilon}
\def\v{\B}
\def\w{\omega}
\def\x{\xi}
\def\y{\eta}
\def\z{\zeta}
\def\D{\Delta}
\def\G{\Gamma}
\def\H{\Theta}
\def\L{\Lambda}
\def\F{\Phi}
\def\P{\Psi}
\def\S{\Sigma}

\def\o{\over}
\newcommand{\gsim}{ \mathop{}_{\textstyle \sim}^{\textstyle >} }
\newcommand{\lsim}{ \mathop{}_{\textstyle \sim}^{\textstyle <} }
\newcommand{\vev}[1]{ \left\langle {#1} \right\rangle }
\newcommand{\bra}[1]{ \langle {#1} | }
\newcommand{\ket}[1]{ | {#1} \rangle }
\newcommand{\EV}{ {\rm eV} }
\newcommand{\KEV}{ {\rm keV} }
\newcommand{\MEV}{ {\rm MeV} }
\newcommand{\GEV}{ {\rm GeV} }
\newcommand{\TEV}{ {\rm TeV} }
\def\diag{\mathop{\rm diag}\nolimits}
\def\Spin{\mathop{\rm Spin}}
\def\SO{\mathop{\rm SO}}
\def\O{\mathop{\rm O}}
\def\SU{\mathop{\rm SU}}
\def\U{\mathop{\rm U}}
\def\Sp{\mathop{\rm Sp}}
\def\SL{\mathop{\rm SL}}
\def\tr{\mathop{\rm tr}}

\def\IJMP{Int.~J.~Mod.~Phys. }
\def\MPL{Mod.~Phys.~Lett. }
\def\NP{Nucl.~Phys. }
\def\PL{Phys.~Lett. }
\def\PR{Phys.~Rev. }
\def\PRL{Phys.~Rev.~Lett. }
\def\PTP{Prog.~Theor.~Phys. }
\def\ZP{Z.~Phys. }
\newcommand{\bear}{\begin{array}}  \newcommand{\eear}{\end{array}}
\newcommand{\bea}{\begin{eqnarray}}  \newcommand{\eea}{\end{eqnarray}}
\newcommand{\beq}{\begin{equation}}  \newcommand{\eeq}{\end{equation}}
\newcommand{\bef}{\begin{figure}}  \newcommand{\eef}{\end{figure}}
\newcommand{\bec}{\begin{center}}  \newcommand{\eec}{\end{center}}
\newcommand{\non}{\nonumber}  \newcommand{\eqn}[1]{\beq {#1}\eeq}
\newcommand{\lmk}{\left(}  \newcommand{\rmk}{\right)}
\newcommand{\lkk}{\left[}  \newcommand{\rkk}{\right]}
\newcommand{\lhk}{\left \{ }  \newcommand{\rhk}{\right \} }
\newcommand{\del}{\partial}  \newcommand{\abs}[1]{\vert{#1}\vert}
\newcommand{\vect}[1]{\mbox{\boldmath${#1}$}}
\newcommand{\bib}{\bibitem} \newcommand{\new}{\newblock}
\newcommand{\la}{\left\langle} \newcommand{\ra}{\right\rangle}
\newcommand{\bfx}{{\bf x}} \newcommand{\bfk}{{\bf k}}
\newcommand{\gtilde} {~ \raisebox{-1ex}{$\stackrel{\textstyle >}{\sim}$} ~} 
\newcommand{\ltilde} {~ \raisebox{-1ex}{$\stackrel{\textstyle <}{\sim}$} ~}
\newcommand{\gtrsim}{ \mathop{}_{\textstyle \sim}^{\textstyle >} }
\newcommand{\lesssim}{ \mathop{}_{\textstyle \sim}^{\textstyle <} }
\newcommand{\ds}{\displaystyle}
\newcommand{\bi}{\bibitem}
\newcommand{\lar}{\leftarrow}
\newcommand{\rar}{\rightarrow}
\newcommand{\lrar}{\leftrightarrow}
\def\Frac#1#2{{\displaystyle\frac{#1}{#2}}}
\def\labelenumi{(\roman{enumi})}
\def\SEC#1{Sec.~\ref{#1}}
\def\FIG#1{Fig.~\ref{#1}}
\def\EQ#1{Eq.~(\ref{#1})}
\def\EQS#1{Eqs.~(\ref{#1})}
\def\lrf#1#2{ \left(\frac{#1}{#2}\right)}
\def\lrfp#1#2#3{ \left(\frac{#1}{#2}\right)^{#3}}
\def\GEV#1{10^{#1}{\rm\,GeV}}
\def\MEV#1{10^{#1}{\rm\,MeV}}
\def\KEV#1{10^{#1}{\rm\,keV}}


\baselineskip 0.7cm

\begin{titlepage}

\begin{flushright}
\hfill DESY 06-149\\
\hfill hep-ph/0611055\\
\hfill November, 2006\\
\end{flushright}

\vskip 1.35cm
\begin{center}
{\large \bf
Spontaneous Non-thermal Leptogenesis in High-scale Inflation Models
}
\vskip 1.2cm
Motoi Endo$^{1}$, Fuminobu Takahashi$^{1}$ 
and T. T. Yanagida$^{2,3}$
\vskip 0.4cm
${}^1${\it Deutsches Elektronen Synchrotron DESY, Notkestrasse 85,\\
22607 Hamburg, Germany}\\
${}^2${\it Department of Physics, University of Tokyo,\\
     Tokyo 113-0033, Japan}\\
${}^3${\it Research Center for the Early Universe, University of Tokyo,\\
     Tokyo 113-0033, Japan}

\vskip 1.5cm

\abstract{ We argue that a non-thermal leptogenesis occurs
spontaneously, without direct couplings of the inflaton with
right-handed neutrinos, in a wide class of high-scale inflation
models such as the chaotic and hybrid inflation. It is only a finite
vacuum expectation value of the inflaton, or more precisely, a linear
term in the K\"ahler potential, that is a prerequisite for the
spontaneous non-thermal leptogenesis. To exemplify how it works, we
show that a chaotic inflation model in supergravity naturally produces
a right amount of baryon asymmetry via the spontaneous non-thermal
leptogenesis.  We also discuss the gravitino production from the
inflaton.  }
\end{center}
\end{titlepage}

\setcounter{page}{2}

\section{Introduction}
\label{sec:1}
The origin of the baryon asymmetry in the universe is one of the most
important issues in the particle cosmology. Among many baryogeneses
proposed so far, particularly interesting is the leptogenesis
scenario~\cite{Fukugita:1986hr}, which may be divided into two
classes, thermal and non-thermal ones, depending on the mechanism to
generate the heavy right-handed neutrinos.  The thermal leptogenesis
is simple, and therefore attractive; it almost automatically takes
place once the cosmic temperature rises so high that the right-handed
neutrinos are thermalized. Since it requires relatively high reheating
temperatures, it is consistent with the big bang nucleosynthesis (BBN)
only for a limited range of the gravitino mass.  The non-thermal
leptogenesis, on the other hand, is capable of producing the baryon
asymmetry at lower reheating temperatures, allowing a broader range of
the gravitino mass. However, since the generation of the right-handed
neutrinos relies on couplings of the inflaton, one may get the
impression that it needs ad hoc model-dependent assumptions in
comparison with the thermal leptogenesis.

Recently there has been much progress concerning the decay processes
of scalar fields such as moduli~\cite{moduli,Dine:2006ii,Endo:2006tf,Endo:2006ix}
and inflaton~\cite{Kawasaki:2006gs,Asaka:2006bv,Endo:2006qk} in
supergravity. In particular, Ref.~\cite{Endo:2006qk} has pointed out
that, if the inflaton has a linear term in the K\"ahler
potential~\footnote{ Note that the K\"{a}hler potential contains a
linear term when the inflaton acquires a finite vacuum expectation
value (VEV), even if the minimal K\"{a}hler potential is assumed from
the beginning.  }, the inflaton couples to all the fields that appear
in the superpotential with gravitational strength~\footnote{
T.T.Y. thanks T. Watari for a useful discussion on the point.
}.  The discovery of such couplings has opened a way to naturally
induce the reheating into the visible sector.  In particular, it
enables the non-thermal leptogenesis~\cite{Asaka:1999yd,
Lazarides:1993sn} to occur more naturally, in the sense that one does
not have to introduce any direct couplings of the inflaton with the
right-handed neutrinos.

In this paper we argue that the non-thermal leptogenesis naturally
occurs in a certain class of high-scale inflation models~\footnote{
The inflaton must be heavy enough since otherwise the reheating
temperature is too low for the leptogenesis to work.  See
Eq.~(\ref{eq:low-Tr}) below.  }, without introducing couplings of the
inflaton with the right-handed neutrinos. The presence of the linear
term in the K\"ahler potential, which may result from the inflaton's
VEV, plays an essential role in our discussion. To illustrate how it
works, we investigate a chaotic inflation model in supergravity and
show that the spontaneous non-thermal leptogenesis actually takes
place and generates a right amount of the baryon asymmetry.  Although
we consider mainly the chaotic inflation model~\footnote{
The recent WMAP three year results have shown that the scalar spectral
index is quite likely to be smaller than unity: $n_s =
0.951^{+0.015}_{-0.019}$~\cite{Spergel:2006hy}.  Among inflation
models consistent with the WMAP results, from the observational point
of view, the chaotic inflation model is interesting, because the
predicted B-mode signal in CMB polarization is likely to be detected
by future observations such as Planck, Clover, and CMBpol.
}, the results derived in the following are rather generic, and can be
applied to any high-scale inflation models.

The paper is organized as follows. In Sec.~\ref{sec:2} we review the
chaotic inflation model in supergravity. In Sec.~\ref{sec:3}, we will
discuss the spontaneous non-thermal leptogenesis.  The last section is
devoted to conclusions. The case without a linear term will also be
briefly discussed in Appendix.

\section{Chaotic Inflation Model in Supergravity}
\label{sec:2}
To construct a successful inflation model in supergravity, which we
assume throughout this paper, there is a well-known
$\eta$-problem~\cite{Kumekawa:1994gx}.  The problem becomes much
severer when constructing a chaotic inflation
model~\cite{Linde:1983gd}, since the exponential prefactor of the
scalar potential practically forbids any scalar fields to take values
beyond the Planck scale, while the inflaton must initially sit at the
point far beyond the Planck scale in the chaotic inflation model. It
was pointed out in Ref.~\cite{Kawasaki:2000yn} that the problem can be
solved by postulating a shift symmetry on the inflaton and the chaotic
inflation in supergravity can be realized in a rather simple set-up.

According to Ref.~\cite{Kawasaki:2000yn}, we assume that the K\"ahler
potential $K(\phi,\phi^\dag)$ is invariant under the shift of $\phi$,
\begin{equation}
  \phi \rightarrow \phi + i\,A,
  \label{eq:shift}
\end{equation}
where $A$ is a dimension-one real parameter, and we adopt the Planck
unit: $M_P = 1$ unless stated otherwise. Thus, the K\"ahler potential
is a function of $\phi + \phi^\dag$; $K(\phi,\phi^\dag) =
K(\phi+\phi^\dag)= c\,(\phi+\phi^\dag) + \frac{1}{2}
(\phi+\phi^\dag)^2 + \cdots$, where $c$ is a real constant and must be
smaller than $O(1)$ for a successful inflation.  We will identify the
imaginary part of $\phi$ with the inflaton field $\varphi \equiv
\sqrt{2} {\rm \,Im}[\phi]$.  Moreover, we introduce a small breaking
term of the shift symmetry in the superpotential in order for the
inflaton $\varphi$ to have a potential:
\begin{equation}
  W(\phi,\psi) = m\,\phi \,\psi, 
  \label{eq:mass}
\end{equation}
where we have introduced a new chiral multiplet $\psi$, and $m \simeq
2\times10^{13}$GeV represents the breaking scale of the shift symmetry
and determines the inflaton mass.

The scalar potential is given by
\bea 
V(\eta, \varphi, \psi) &=& m^2 e^{K} \left[|\psi|^2 \left(1+2
\left(\eta + \frac{c}{\sqrt{2}}\right)\eta + \left(\eta +
\frac{c}{\sqrt{2}}\right)^2(\eta^2+\varphi^2)\right)\right.\non\\ &&
\left.~~~~~~~~~+\frac{1}{2}(\eta^2+\varphi^2)(1-|\psi|^2+|\psi|^4)\right]
\eea
with \beq K \;=\; \left(\eta +
\frac{c}{\sqrt{2}}\right)^2-\frac{c^2}{2}+|\psi|^2, \eeq where we have
assumed the minimal K\"ahler potential for $\psi$, and defined $\eta
\equiv \sqrt{2} {\rm\,Re}[\phi]$.  Note that $\eta$ and $\psi$ cannot
be larger than the Planck scale, due to the prefactor $e^K$.  On the
other hand, $\varphi$ can be larger than the Planck scale, since
$\varphi$ does not appear in $K$.  For $\varphi \gg 1$, $\eta$
acquires the mass comparable to the Hubble parameter and quickly
settles down to the minimum, $\eta \simeq -c/\sqrt{2}$. Then the
scalar potential during inflation is given by
\beq
V(\eta,\varphi,\psi) \;\simeq\; \frac{1}{2}m^2 \varphi^2 + m^2 |\psi|^2.
\eeq
For $\varphi \gg 1$ and $|\psi| < 1$, the $\varphi$ field dominates
the potential and the chaotic inflation takes place (for details see
Ref.~\cite{Kawasaki:2000yn}).

When $\varphi \simeq \sqrt{2}$, the slow-roll condition breaks down
and the inflaton $\varphi$ starts to oscillate. The potential minimum
after inflation is located at $\varphi=\eta=0$ and $\psi=0$ in the
SUSY limit.  Once we take account of the SUSY breaking~\footnote{
For a broad range of the gravitino mass from $O(10)$eV to $O(100)$TeV,
the Hubble parameter at the reheating must be smaller than the
gravitino mass. Otherwise,  too many gravitinos would be produced by
particle scatterings in thermal plasma~\cite{Kawasaki:2006gs} unless 
there is late-time entropy production~\cite{Lyth:1995ka}.
Therefore one needs to take account
of the SUSY breaking when considering the inflaton decay.
}, the minimum slightly changes, although the shift is so tiny that the
following discussion is not affected. The other important effect is
the mixing between $\phi$ and $\psi^\dag$ induced by the SUSY
breaking. As discussed in Ref.~\cite{Kawasaki:2006gs}, $\phi$ and
$\psi^\dag$ almost maximally mix with each other to form the mass
eigenstates:
\beq
\varphi_\pm \; \equiv \; \frac{\phi \pm \psi^\dag}{\sqrt{2}}.
\eeq
Therefore we will consider the decay processes of $\varphi_\pm$ instead of
$\varphi$, $\eta$ and $\psi$.

\section{Spontaneous Non-thermal Leptogenesis}
\label{sec:3}

The decay processes on which we focus our attention are those induced
by the presence of the linear term in the K\"ahler potential: $\delta
K = c (\phi+\phi^\dag)$.  Such a linear term is expected to exist with
a coefficient of order unity, $c \lesssim O(1)$~\footnote{ $c \lesssim
O(1)$ is required for a successful inflation~\cite{Kawasaki:2000yn}.
}, since it is consistent with the shift symmetry~(\ref{eq:shift}).
Although the inflaton may have direct couplings with matter fields in
the K\"ahler potential and the superpotential, we assume that such
couplings are suppressed.

The inflaton can decay into all the fields that appear in the
superpotential through the linear term with gravitational
strength~\cite{Endo:2006qk}, if it is kinematically allowed, and we
take up the following two important processes above all.  First let us
consider the inflaton decay through the top Yukawa coupling:
\beq
W \;=\; Y_t \,T Q H_u,
\eeq
where $Y_t$ is the top Yukawa coupling, and $T$, $Q$, and $H_u$ are
the chiral supermultiplets of the right-handed top quark and
left-handed quark doublet of the third generation, and up-type Higgs,
respectively. The partial decay rate of the inflaton through the top
Yukawa coupling is~\cite{Endo:2006qk}
\beq
\label{eq:rate-th-top}
\Gamma_T \;\simeq\; \frac{3}{128 \pi^3} |Y_t|^2 \lrfp{c}{\sqrt{2}}{2} 
\frac{m^3}{M_P^2},
\eeq
where we have taken account of the mixing mentioned in the end of the
previous section. Due to the decay process via the top Yukawa
coupling, the reheating temperature $T_R$ is bounded below. We define
the reheating temperature as
\beq
\label{eq:def-Tr}
T_R \;\equiv\; \lrfp{\pi^2 g_*}{10}{-\frac{1}{4}} \sqrt{\Gamma_\varphi M_P},
\eeq
where $g_* $ counts the relativistic degrees of freedom, and
$\Gamma_\varphi$ denotes the total decay rate of the inflaton.  Using
$\Gamma_\varphi \geq \Gamma_T$, we obtain
\beq
\label{eq:low-Tr}
T_R \;\gtrsim \; 2 \times 10^8 {\rm \,GeV}\, |c| 
\lrfp{m}{2\times10^{13}{\rm GeV}}{\frac{3}{2}},
\eeq
where we have substituted $g_* = 228.75$ and $Y_t \simeq 0.6$~\footnote{
Here we have estimated the top Yukawa coupling at an energy 
scale of the inflaton mass $\simeq O(10^{13})$GeV.
}, and the
inequality is saturated if $\Gamma_\varphi \simeq \Gamma_T$.  It is
quite striking that the inflaton can decay into the visible sector
even without direct couplings in the K\"ahler potential or the
superpotential.

The other important process we consider is the decay into the
right-handed (s)neutrinos thorough large Majorana mass terms:
\beq
W \;=\; \frac{M_i}{2}  N_i N_i,
\eeq
where $i = 1,2,3$ is the family index.  We consider the inflaton decay
into the lightest right-handed (s)neutrino $N_1$ for simplicity,
assuming that the decay into the heavier ones, $N_2$ and $N_3$, are
kinematically forbidden.  We drop the family index in the following.
The partial decay rate of the inflaton into the right-handed
(s)neutrinos is [cf.~\cite{Endo:2006qk}]
\beq
\label{eq:decay-N}
\Gamma_N \;\simeq\; \frac{1}{16 \pi} \lrfp{c}{\sqrt{2}}{2} 
\frac{m M^2 }{M_P^2} \sqrt{1-\frac{4M^2}{m^2}},
\eeq
where we have taken account of both the decay into the right-handed
neutrinos and that into the right-handed sneutrinos. Since $\Gamma_N$
is proportional to $M^2$, it is much smaller than $\Gamma_T$ for $M
\ll m$, but it can be comparable to or even larger than $\Gamma_T$ for
$M = O(10^{12})\,$GeV.

The lepton asymmetry can be produced by the decay of the right-handed
(s)neutrinos, if $CP$ is violated in the neutrino Yukawa
matrix~\cite{Fukugita:1986hr}.  The resultant lepton asymmetry is
given by
\beq
\frac{n_L}{s} \;\simeq\; \frac{3}{2} \epsilon\, B_N \frac{T_R}{m},
\eeq
where $B_N \equiv \Gamma_N/\Gamma_\varphi$ denotes the branching ratio
of the inflaton decay into the (s)neutrinos.  The asymmetry parameter
$\epsilon$ is given by~\cite{Fukugita:1986hr,Covi:1996wh}
\beq
\epsilon \simeq 2.0 \times 10^{-10} \lrf{M}{10^6{\rm GeV}} 
\lrf{m_{\nu_3}}{0.05{\rm eV}} \delta_{\rm eff},
\eeq
where $m_{\nu_3}$ is the heaviest neutrino mass and $ \delta_{\rm eff}
\leq 1$ represents the effective $CP$-violating phase. The baryon
asymmetry is obtained via the sphaleron
effect:~\cite{Khlebnikov:1988sr}
\beq
\frac{n_B}{s} \;=\; -\frac{8}{23} \frac{n_L}{s}. 
\eeq
Using  the above relations, we obtain the right amount of baryon asymmetry,
\beq
\label{eq:nbs}
\frac{n_B}{s} \;\simeq\; 2 \times 10^{-10}\, |c| \lrfp{M}{10^{12}{\rm GeV}}{3} 
\lrfp{m}{2\times10^{13}{\rm GeV}}{-\frac{3}{2}} \lrf{m_{\nu_3}}{0.05{\rm eV}} \delta_{\rm eff},
\eeq
where we have approximated that the inequality (\ref{eq:low-Tr}) is
saturated, assuming $M \ll m$.

\begin{figure}[t!]
\begin{center}
\includegraphics[width=10cm]{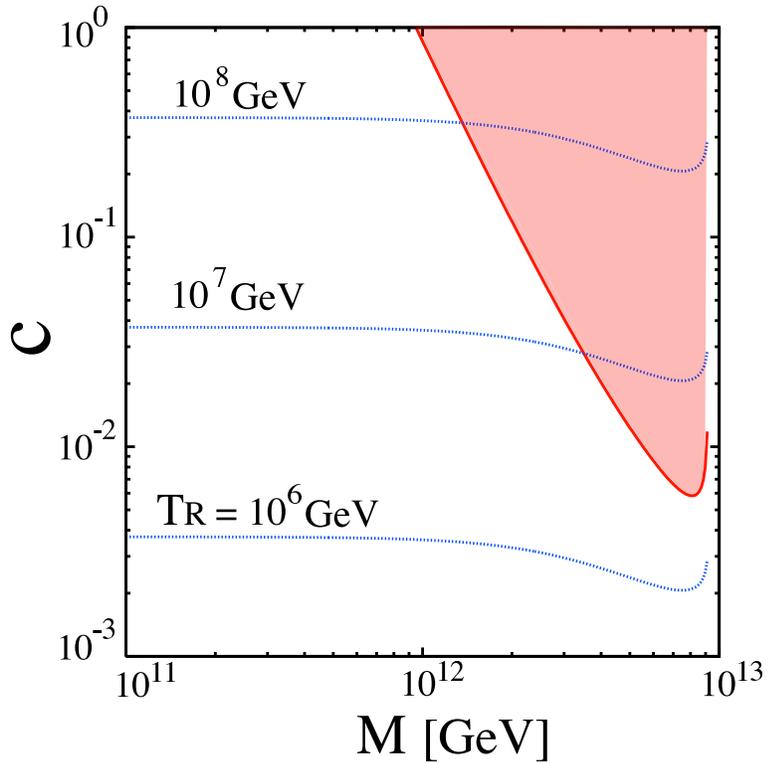}
\caption{Contours of the reheating temperatures $T_R$, denoted by
dotted (blue) lines.  Taking account of the unknown $CP$ phase 
$\delta_{\rm eff} \leq 1$, the shaded region above the solid (red) line 
can explain the present baryon asymmetry.
We set $g_* \;=\; 228.75$, $Y_t \;=\; 0.6$ and
$m_{\nu_3} = 0.05\,$eV. }
\label{fig:tr-low}
\end{center}
\end{figure}

In Fig.~\ref{fig:tr-low}, we plot the contours of the reheating
temperatures as a function of $c$ and $M$. Note that the total decay
rate $\Gamma_\varphi$ is given by $\Gamma_T + \Gamma_N$, as long as $M
< m/2$.  In addition, we show the region where the non-thermal
leptogenesis can explain the baryon asymmetry measured by
WMAP~\cite{Spergel:2006hy}:
\beq
\left.\frac{n_B}{s}\right|_{\rm WMAP} \;=\; 8.7^{+0.3}_{-0.4} \times 10^{-11}.
\eeq
From Fig.~\ref{fig:tr-low}, one can see that the coefficient of the
linear term in the K\"ahler potential, the reheating temperature and
the (lightest) right-handed neutrino mass must be in the following
ranges:
\bea
\label{bound-on-c}
6 \times 10^{-3} \; \lesssim & c &\lesssim\; 1,\\
\label{bound-on-Tr}
3 \times 10^6 {\rm\,GeV}\; \lesssim&T_R &\lesssim\; 3 \times 10^8 {\rm\,GeV},\\
\label{bound-on-M}
10^{12} {\rm\,GeV} \;\lesssim &M& \lesssim \;10^{13}{\rm \,GeV},
\eea
for the successful non-thermal leptogenesis.  The parameter ranges
shown above are the prediction of our non-thermal leptogenesis
scenario in the chaotic inflation model. Note that such a large
right-handed neutrino mass may induce large lepton flavor violating
signals~\cite{Tobe:2003nx} at a detectable level by MEG~\cite{meg}.

The reheating temperature in the above range (\ref{bound-on-Tr})
constrains the gravitino mass to satisfy the bounds from the gravitino
problem~\cite{gravitino-problem}~\footnote{
The recent results are given in Refs.~\cite{Kohri:2001jx} for the
unstable gravitino and in Ref.~\cite{Moroi:1993mb} for the stable one.
See Ref.~\cite{Kawasaki:2006gs} for the summarized results.
}. The abundance of the gravitino produced by thermal scatterings 
is approximated by~\cite{Bolz:2000fu,Kawasaki:2004yh}
\begin{eqnarray}
    \label{eq:Yx-new}
    Y_{3/2} &\simeq& 
    1.9 \times 10^{-12}\left[ 1+ 
    \left(\frac{m_{\tilde{g}_3}^2}{3m_{3/2}^2}\right)\right]
    \left( \frac{T_{\rm R}}{10^{10}\ {\rm GeV}} \right),
\end{eqnarray}
where $m_{\tilde{g}_3}$ is the gluino mass evaluated at $T=T_R$, and
we have dropped the logarithmic corrections for simplicity.  If the
gravitino is light, it may be the lightest SUSY particle (LSP) and
therefore stable. The bounds on $T_R$ then come from the requirement
that the gravitino abundance should not exceed the present dark matter
(DM) abundance:
\bea
\label{eq:stable-g2}
  T_R \; \lesssim \; \left\{
  \bear{ll}
  O(100)~{\rm GeV} &
  {\rm for}~~~m_{3/2}\simeq 10^{-2} - 10^{2}~{\rm keV}\\
\ds{3\times 10^7~{\rm GeV} 
   \lrfp{m_{\tilde{g}_3}}{500{\rm\,GeV}}{-2}   \left(\frac{m_{3/2}}{1{\rm \,GeV}}\right)}
   &{\rm for}~~~m_{3/2}\simeq    10^{-4} - 10~{\rm GeV}
\eear  \right. .
\eea
 On the other hand, if the gravitino is unstable,
 BBN puts severe constraints on $T_R$:
\bea
\label{eq:rtemp_0}
   T_R\; \lesssim\; \left\{ \bear{ll} (1-4)\times 10^6~{\rm GeV} &{\rm
   for} ~~ m_{3/2} \simeq 0.1 - 0.2~{\rm TeV} \\ 3 \times 10^{5} - 4
   \times 10^{6}~{\rm GeV} &{\rm for} ~~ m_{3/2} \simeq 0.2 - 2~{\rm
   TeV} \\ 5\times 10^{5} - 1\times 10^{8}~{\rm GeV} & {\rm for} ~~
   m_{3/2} \simeq 2 - 10~{\rm TeV} \\ (3 -10)\times 10^9~{\rm GeV}
   &{\rm for} ~~ m_{3/2} \simeq 10 - 30~{\rm TeV} \eear \right., \eea
for the hadronic branch $B_h =1$. For smaller $B_h$, the bounds are
relaxed to some extent.  For the heavy gravitino of mass $\gtrsim 30$
TeV, no stringent constraints are obtained from BBN.  However, another
constraint comes from the abundance of the LSP produced by the
gravitino decay:
\beq
T_R \;\lesssim\; 9.3 \times 10^{9} \lrfp{m_{3/2}}{100{\rm\,TeV}}{-1} {\rm GeV},
\label{eq:const-from-winolsp}
\eeq
where we have assumed the anomaly-mediated SUSY breaking
(AMSB)~\cite{AMSB} with the wino LSP, and used the relation
$m_{\tilde{W}} \simeq 2.7 \times 10^{-3} m_{3/2}$~\footnote{
This relation may be changed by radiative corrections~\cite{Giudice:1998xp}.
}.  Thus, the allowed range for $T_R$ (\ref{bound-on-Tr}) satisfies
the bounds shown above, if the gravitino mass $m_{3/2}$ lies roughly
in the following ranges:~\footnote{
Taking account of the decay of the next-to-lightest SUSY 
particle~\cite{NLSP-decay,stau-bound},
the upper bound of the case (ii) may become severer.
}
\bea
&&{\rm (i)}\, m_{3/2} \;\lesssim\; 10\,{\rm eV},\non\\
\label{eq:mg}
&&{\rm (ii)}\, 10{\rm\, MeV} \;\lesssim \;m_{3/2}\;\lesssim\; 10{\rm\, GeV},\\
&&{\rm (iii)}\, m_{3/2}\; \gtrsim \;10{\rm \,TeV}.\non
\eea
For the case (i), the bound on $T_R$ does not exist, since the
gravitino is so light that the contribution to the dark matter is
negligibly small.  For the cases (ii) and (iii), there are further
constraints on the SUSY breaking sector from the non-thermal gravitino
production by the inflaton decay~\cite{Kawasaki:2006gs}, as will be
discussed below.

Let us now discuss the gravitino pair production from the
inflaton~\cite{Kawasaki:2006gs}.  Even if the inflaton does not couple
to the SUSY breaking field $z$, the gravitino pair production still
occurs, and it is efficient especially if the SUSY breaking field $z$
has a large mass $m_z$~\cite{Dine:2006ii,Endo:2006tf} as in the
dynamical SUSY breaking (DSB) scenario~\cite{Izawa:1996pk}.  Assuming
$m_z < m$ and no direct couplings between the inflaton and $z$, the
gravitino production rate is~\cite{Endo:2006tf}
\beq
\Gamma_{3/2}^{\rm (pair)} \;\simeq\; \frac{|c|^2}{64 \pi} \frac{m^3}{M_P^2} \lrfp{m_z}{m}{4}.
\eeq
The resultant gravitino abundance is
\bea
Y_{3/2}^{\rm (pair)} &\simeq& 
5\times10^{-4} |c| \lrfp{m}{2\times10^{13}{\rm GeV}}{\frac{1}{2}}  \lrfp{m_z}{m}{4},
\eea
where we have used (\ref{eq:low-Tr}) assuming $\Gamma_\varphi \simeq
\Gamma_T$.  Note that, since the gravitino abundance is proportional
to the fourth power of the mass ratio $(m_z/m)$, it can be suppressed
if $m_z$ is much smaller than the inflaton mass $m$. For instance, in
the case of AMSB with the wino-like LSP, the bound on $Y_{3/2}$ is
obtained from (\ref{eq:Yx-new}) and (\ref{eq:const-from-winolsp}):
\beq
\label{Yamsb}
Y_{3/2} \;\lesssim\; 2 \times 10^{-12} \lrf{100{\rm TeV}}{m_{3/2}}.
\eeq
Using this bound, we obtain a mild constraint on $m_z$,
\beq
\label{eq:amsb}
m_z  \; \lesssim\; 2 \times 10^{11} {\rm\, GeV}\, |c|^{-1/4} \lrfp{m}{2\times10^{13}{\rm GeV}}{7/8} 
		\lrfp{m_{3/2}}{100{\rm TeV}}{-1/4}~~~~{\rm for~~AMSB}.
\eeq
On the other hand, for the lighter gravitino ($\lesssim 10{\rm\,GeV}$)
as in the gauge-mediated SUSY breaking (GMSB)~\cite{GMSB} which
corresponds to the case (ii), the gravitino should be the LSP. The
bound on $Y_{3/2}$ reads
\beq
\label{Ygmsb}
Y_{3/2}\; \lesssim\; 5 \times 10^{-10} \lrf{1{\rm GeV}}{m_{3/2}}.
\eeq 
Then the constraint on $m_z$ becomes
\beq
\label{eq:gmsb}
m_z \; \lesssim\; 6 \times 10^{11} {\rm\, GeV}\, |c|^{-1/4}
		\lrfp{m}{2\times10^{13}{\rm GeV}}{7/8}
		\lrfp{m_{3/2}}{1{\rm GeV}}{-1/4}~~~~{\rm for~~GMSB}.
		\eeq
Since $m_z$ is likely smaller for the lighter gravitinos, the bound (\ref{eq:gmsb})
is much milder than (\ref{eq:amsb}).
Thus, although the gravitino pair
production puts some constraints on the SUSY breaking models, they are
not so severe~\footnote{
Note that the gravity pair production severely constrains the
gravity-mediated SUSY breaking models that contain a singlet with
non-zero $F$-term, unless the inflaton has some unbroken symmetries in
vacuum~\cite{Kawasaki:2006gs}.
}.

So far we have focused on the two decay processes among those induced
by the linear term in the K\"ahler potential. As pointed out in
Ref.~\cite{Endo:2006qk}, the gravitinos may be directly produced in a
similar way, if the DSB sector has Yukawa couplings in the
superpotential, and if the decay into the DSB sector is kinematically
allowed.  The gravitino production rate is expressed by
\beq
\label{gravitino-rate}
\Gamma_{3/2} \;=\; 
\frac{\xi}{1536 \pi^3}
\lrfp{c}{\sqrt{2}}{2}
\frac{m^3}{M_P^2},
\eeq
where $\xi$ is determined by the decay processes; the degrees of
freedom of the decay products, the decay chains, the coupling
constants of the Yukawa interactions in the SUSY breaking sector, and
form factors of the hidden mesons and/or baryons. Although the
constant $\xi$ strongly depends on the models, if all the Yukawa
couplings are of order unity, $\xi$ is expected to be $O(1)$ or
larger.  The gravitino abundance is then
\bea
Y_{3/2} &=& 
5 \times 10^{-7}\, \xi\, |c|
\lrfp{m}{2\times10^{13}{\rm GeV}}{\frac{1}{2}}.
\label{gravitinoY}
\eea
To avoid the gravitino overproduction, $\xi$ must be small enough. Using the
upper bounds on $Y_{3/2}$, (\ref{Yamsb}) and (\ref{Ygmsb}), we obtain
\beq
\label{eq:xi}
\xi \;\lesssim\;
\left\{
\bear{ll}
\ds{4\times 10^{-6} |c|^{-1}  \lrfp{m}{2\times10^{13}{\rm GeV}}{-\frac{1}{2}}
 \lrfp{m_{3/2}}{100{\rm TeV}}{-1}}&{\rm for~~AMSB},\\
\ds{1 \times 10^{-3} |c|^{-1}  \lrfp{m}{2\times10^{13}{\rm GeV}}{-\frac{1}{2}} 
\lrfp{m_{3/2}}{1{\rm GeV}}{-1}}&{\rm for~~GMSB}.
\eear
\right.
\eeq
Thus, the Yukawa couplings of order unity in the DSB sector may not be
allowed especially in the case of the AMSB scenario~\footnote{
It is marginally possible that the inflaton decay into the DSB sector
is kinematically forbidden for $m_{3/2} \sim O(100) {\rm\,TeV}$.  Then
the constraint on $\xi$ is not applied.
}.  Note that there are the DSB models with no
superpotential~\cite{Affleck:1983vc,Affleck:1984uz,Murayama:1995ng}
(i.e., $\xi = 0$), which evade the gravitino overproduction problem
mentioned here~\footnote{
The constraint on $\xi$ may be applicable even in this case, since the
inflaton mass is close to the DSB scale in the AMSB scenario and there
can be non-perturbative effects inducing the effective coupling
$\xi_{\rm eff}$ for composite states.
}.

We have paid our attention so far to the baryon asymmetry generated
via leptogenesis.  For successful cosmology after inflation, a right
amount of DM in addition to the baryon asymmetry must be
generated. For the case (i) shown in (\ref{eq:mg}), the gravitino
should be the LSP, but it is too light to be the dominant component of
DM. Therefore we need to introduce e.g. the axion to account for
DM. In the case (ii), the gravitino can be DM, and it comes both from
the thermal production and from the inflaton decay.  In the case
(iii), the LSP is likely the wino, which is also generated from the
thermal production and the decay of the gravitinos.

\section{Discussion and Conclusions}
\label{sec:4}
As we have seen in the previous section, the linear term in the
K\"ahler potential plays an essential role in the spontaneous
non-thermal leptogenesis scenario. Once the inflaton acquires such a
linear term, the non-thermal leptogenesis follows automatically.
Although we have considered the chaotic inflation model to illustrate
the mechanism, it can be applied to any high-scale inflation models in
which the inflaton acquires a finite VEV, by replacing $c/\sqrt{2}$
with the VEV and $m$ with the inflaton mass of the model under
consideration.  Of course, the allowed ranges of the parameters and
the constraints on the SUSY breaking sector from the gravitino
overproduction problem depend on the inflation model.

For instance, in the case of the hybrid inflation
model~\cite{Copeland:1994vg}, the waterfall field obtains a finite VEV
$\la \phi \ra$, which is related to the inflaton mass $m_\phi$ as
$m_\phi \;=\; \sqrt{2} \lambda \la \phi \ra$. Here $\lambda$ is a
Yukawa coupling between the inflaton and the waterfall fields, and it
takes a value ranging from $10^{-5}$ to $10^{-1}$~\cite{Bastero-Gil:2006cm}. 
Let us concentrate on $\lambda \gtrsim 10^{-3}$, since otherwise the scalar spectral
index would become close to unity, which is disfavored by the recent
WMAP data~\cite{Spergel:2006hy}. Then the VEV and the mass are given
by $\la \phi \ra \;\simeq\; 2 \times 10^{-3}$ and $m_\phi \;\simeq\; 3
\times 10^{-3} \lambda$.  Replacing $c/\sqrt{2}$ with $\la \phi \ra$
and $m$ with $m_\phi$ in Eq.~(\ref{eq:nbs}), we obtain the baryon
asymmetry produced by the spontaneous non-thermal leptogenesis in the
hybrid inflation model:
\beq \left.\frac{n_B}{s}\right|_{\rm hybrid} \;\simeq\; 9 \times
10^{-11}\, \lrfp{\lambda}{10^{-3}}{3/2} \lrfp{M}{m_\phi/2}{3}
\lrf{m_{\nu_3}}{0.05{\rm eV}} \delta_{\rm eff}.  \eeq
Due to the smaller VEV $\la \phi \ra$ compared to $c$ in the chaotic
inflation model, a right amount of the baryon asymmetry is generated
only if $M$ takes a value close (but not too close) to the upper bound
$m_\phi/2$. Thus the spontaneous non-thermal leptogenesis works in the
hybrid inflation model too. The constraints on the SUSY breaking
sector can be similarly read from (\ref{eq:amsb}), (\ref{eq:gmsb}) and
(\ref{eq:xi}) by replacing $c/\sqrt{2}$ with $\la \phi \ra$ and $m$
with $m_\phi$, and they are more or less similar.

In this paper we have investigated the reheating processes of the
chaotic inflation in supergravity, paying particular attention to the
decay processes induced by the linear term in the K\"ahler potential.
We have found that a successful non-thermal leptogenesis takes place
spontaneously, without direct couplings with the right-handed
neutrinos.  It requires the parameters such as $c$, $T_R$, $M$, and
$m_{3/2}$ to be in certain ranges (see (\ref{bound-on-c}),
(\ref{bound-on-Tr}), (\ref{bound-on-M}) and (\ref{eq:mg})).  In
particular, the gravitino mass either heavier or lighter than the weak
scale is allowed, while the gravitino of a mass $O(1)$TeV encounters
cosmological difficulties.  Further, the gravitino production from the
inflaton decay can be avoided if the SUSY breaking sector satisfies
some constraints, which are not so severe.  The chaotic inflation
model we have investigated is therefore cosmologically viable, in the
sense that it can naturally cause the leptogenesis avoiding the
gravitino overproduction problem.

Lastly we should stress again that the spontaneous non-thermal
leptogenesis scenario we have proposed in this paper can be applied to
any high-scale inflation models if the inflaton has a finite VEV.  The
scenario is quite unique in that the non-thermal leptogenesis
automatically occurs, and that it predicts several important
parameters such as the reheating temperature, the right-handed
neutrino mass and the gravitino mass, which may be probed in future
experiments/observations.

\vspace{5mm}

\section*{Acknowledgments}
 The work of T.T.Y. has been supported in part by a Humboldt Research Award.


\vspace{5mm}

\appendix
\section{Case without a linear term}
\label{sec:A-1}

Let us here briefly discuss the case without a linear term in K\"ahler
potential.  This can be realized by imposing a discrete symmetry on
the inflaton potential. Since the linear term is suppressed by the
discrete symmetry, the decay modes discussed in the text are ineffective. In
particular, the non-thermal production of the gravitinos from the
inflaton decay does not occur.

The reheating of the inflaton may occur through the couplings with (A)
Higgs fields or (B) right-handed neutrinos.  In the case (A), we
introduce the following coupling:
\beq
\label{eq:inflaton-higgs}
W_{\rm int} \;=\; h\, \phi H_u H_d,
\eeq
where $h$ is a real numerical coefficient and is naturally small, $h =
O(10^{-5})$, since it breaks the shift symmetry (\ref{eq:shift}).  We
show the charges of $U(1)_R$ and $Z_2$ symmetries on the inflaton and
the standard-model fields in Table \ref{tab1}~\footnote{The smallness
of the $\mu$-term can be understood as a tiny breaking of the $Z_2$
symmetry.}.  The decay rate of the inflaton via the coupling
Eq.~(\ref{eq:inflaton-higgs}) is
\beq
\label{rate}
\Gamma_\varphi^{(A)} \;\simeq\; \frac{1}{4 \pi} \lrfp{h}{\sqrt{2}}{2} m.
\eeq
The reheating temperature is given by
\beq
\label{tr}
T_R \;\simeq\; 3 \times 10^9{\rm GeV} \lrf{h}{10^{-5}}
\lrfp{m}{2\times 10^{13} {\rm GeV}}{\frac{1}{2}},
\eeq
which is high enough for thermal leptogenesis to work.  In fact, for
such high reheating temperatures, the right-handed neutrino with a
mass $M \lesssim T_R$ reaches thermal equilibrium, and the baryon
asymmetry is generated via thermal
leptogenesis:~\cite{Fukugita:1986hr,Covi:1996wh}
\beq
\frac{n_B}{s} \;\simeq\; 3 \times 10^{-11}\, \lrf{\kappa}{0.1} \lrf{M}{10^{9}{\rm GeV}}
 \lrf{m_{\nu_3}}{0.05{\rm eV}} \delta_{\rm eff},
\eeq
where $\kappa$ denotes a suppression factor from the wash-out effect.
In this case, the gravitino mass should be larger than $O(10)$TeV to
satisfy the bounds from the gravitino problem (see (\ref{eq:rtemp_0})).

In the case (B) we consider the following  interaction:
\beq
W_{\rm int} \;=\; \frac{k}{2}\, \phi N N,
\eeq
where $k$ is a real numerical coefficient of $O(10^{-5})$ since it
breaks the shift symmetry.  The charge assignments are shown in Table
\ref{tab2}.  Here we promote $Z_2$ symmetry to $Z_4$ symmetry to
accommodate the inflaton coupling with the right-handed neutrinos.
The $Z_4$ symmetry may be identified with the subgroup of
$U(1)_{B-L}$.  The right-handed neutrino masses break the $Z_4$
symmetry down to $Z_2$ symmetry~\footnote{ The right-handed neutrino
mass term induces the linear term of $\phi$ in the K\"ahler potential
with a coefficient $c \sim O(10^{-2} M)$ at one-loop level.  However,
taking account of $M \lesssim 10^{15}{\rm\,GeV}$, the coefficient $c$
is so small that the inflaton decay through the induced linear term
can be neglected.  }.  The decay rate is given by
\beq
\Gamma_\varphi^{(B)} \;\simeq\; \frac{1}{16\pi} \lrfp{k}{\sqrt{2}}{2} m,
\eeq
leading to the reheating temperature
\beq
\label{tr2}
T_R \;\simeq\; 2 \times 10^9{\rm GeV} \lrf{k}{10^{-5}} \lrfp{m}{2\times 10^{13} {\rm GeV}}{\frac{1}{2}}.
\eeq
The non-thermal leptogenesis occurs in this case. Assuming that the
reheating occurs mainly through the decay into the right-handed
neutrinos, the baryon asymmetry is given by
\beq
\frac{n_B}{s} \;\simeq\; 1 \times 10^{-10}\,\lrf{k}{10^{-5}} \lrf{M}{10^{10}{\rm GeV}} \lrfp{m}{2\times 10^{13} {\rm GeV}}{-\frac{1}{2}}
 \lrf{m_{\nu_3}}{0.05{\rm eV}} \delta_{\rm eff}.
\eeq
Note that a right amount of the baryon asymmetry is generated for $k
\sim 10^{-8}$ and $M \sim 10^{13}$GeV, corresponding to $T_R \sim
10^{6}$GeV being marginally compatible with the gravitino mass of
$O(1)$TeV (see (\ref{eq:rtemp_0}))~\footnote{
However, there is the severe Polonyi problem~\cite{Ibe:2006am} in this case, 
which cannot be solved in our framework.
}.

Thus, if the inflaton has a discrete symmetry and the linear term in
the K\"ahler potential is suppressed, either thermal or non-thermal
leptogenesis is possible depending on the coupling with the matter
fields. In particular, it should be noted that the dangerous decay
processes such as the gravitino pair production and the decay into the
SUSY breaking sector are suppressed by the discrete
symmetry. Therefore there is no gravitino overproduction problem in
this case.

\begin{table}[t]
\begin{center}
\begin{tabular}{|c|c|c|c|c|c|c|c|}
\hline
\hline
                   &$\phi$ & $\psi$ & $H_u$ & $H_d$ & ${\bf 5}^*$ & ${\bf 10}$&$N$ \\ \hline
$U(1)_R$ & $0$     & $2$     & $4/5$      & $6/5$      & $1/5$   &  $3/5$ &$1$ \\ \hline
$Z_2$       & $-$      &  $-$     &  $+$     & $-$       &  $-$    &   $+$& $-$        \\
\hline\hline
\end{tabular}
\end{center}
\caption{The charges of $U(1)_R$ and $Z_2$ symmetries in the case (A).}
\label{tab1}
\end{table}
\begin{table}[t]
\begin{center}
\begin{tabular}{|c|c|c|c|c|c|c|c|}
\hline
\hline
                   &$\phi$ & $\psi$  &$H_u$ & $H_d$ & ${\bf 5}^*$ & ${\bf 10}$ &$N$\\ \hline
$U(1)_R$ & $0$     & $2$        &$0$      & $0$      & $1$          &  $1$ &   1 \\ \hline
$Z_4$       & $2$      &  $2$       &$2$     & $2$      & $1$          &   $1$ &   1 \\
\hline\hline
\end{tabular}
\end{center}
\caption{The charges of $U(1)_R$ and $Z_4$ symmetries in the case (B).}
\label{tab2}
\end{table}

\clearpage

\end{document}